\documentclass[12pt]{article}
\title{Restrictions on the Hadronic Contribution to the Muon
$(g-2)$-factor\\
and to the Pion Electromagnetic Formfactor from Analytical Properties\\
of the Pion Electromagnetic Formfactor}
\author{B.V.Geshkenbein\\
Institute for Theoretical and Experimental Physics\\
RU-117259 Moscow, Russia,\\
E-mail: geshken@vxitep.itep.ru}
\date{}
\begin{document}
\maketitle
%\vspace{3cm}
\centerline{\bf{Abstract}}

\vspace{5mm}
We investigate restrictions on the hadronic contribution to the muon $(g-2)$
factor and to the pion electromagnetic formfactor from the analytical
properties of the pion formfactor and the experimental data of the pion
formfactor in the space-like [1] region. The values of the pion formfactor
of [1] have been improved (see Table 1) and the values of the pion
formfactor have been calculated at 20 additional points (see Table 2).

\newpage
{\bf  1. Introduction}

\vspace{3mm}
The aim of this work is to obtain  restrictions on the hadronic
vacuum-polarization contributions to the anomalous magnetic moment
$a^F_{\mu}$ and to the pion electromagnetic formfactor (f.f.) from the
analytical properties of the pion formfactor and from the measurements of
the pion formfactor in the space-like region at 45 points $^{[1]}$.

The contribution of the pion electromagnetic formfactor to the anomalous
magnetic moment can be written as $^{[2]}$

%1
\begin{equation}
a^F_{\mu} = \frac{\alpha^2}{12 \pi^2} \int \limits^{\infty} _{4m^2_{\pi}}
\frac{ds}{s} K(s) \Biggl (1 - \frac{4m^2_{\pi}}{s} \Biggr )^{3/2} \mid
F_{\pi} (s) \mid ^2
\end{equation}
where

$$
K(s) = x^2(1 - x^2/2) + (1 +x)^2 (1 - x^{-2})\Biggl [ln (1 + x) - x + x^2/2
\Biggr ] + $$

%2
\begin{equation}
+ \frac{1 + x}{1 - x} x^2 lnx, ~~~~~ x = \frac{1 - (1 -
4m^2_{\pi}/s)^{1/2}}{1 + (1 - 4m^2_{\pi}/s)^{1/2}}
\end{equation}
$m_{\mu}$ is the muon mass, $m_{\pi}$ is the pion mass.

Function $F_{\pi}(s)$ has the following analytical properties:\\
1. $F_{\pi}(s)$ is an analytical function with the cut $[4 m^2_{\pi},
\infty]$.\\
2. $F_{\pi}(s)$ is a real function on the real axis $[-\infty, 4m^2_{\pi}
 ]$.\\
3.$F_{\pi}(0) = 1$.

The article is organized as follows:

In Sec.2 we find the minimum of $a^F_{\mu}$ from analytical properties
$F_{\pi}(s)$ only.  In Sec.3 the formula for the minimum $a^F_{\mu}$
is obtained for the ease when $F_{\pi}(s)$ has given values $b_k$ at given
points $s_k$

%3
\begin{equation}
F_{\pi} (s_k) = b_k, ~~~~~ k = 0,1, ... N
\end{equation}

In Sec.4 we show that if $F_{\pi}(s_k)$ is exactly equal to the central
value of $F_{\pi}(s_k)$  for 45 points from  $^{[1]}$,
the minimal value of $a^h_{\mu}$ will be by $10^{100}$ times larger than the
conventional value $a^F_{\mu} \approx 4.10^{-8}$ (see Table 1). Sec.5 shows
the way to resolve the contradiction of Sec.4. It is necessary to allow
$F_{\pi}(s_k)$ to vary within the limits of experimental errors. In this
case min $a^F_{\mu} \approx 10^{-8}$.

In Sec.6 it is shown that the use of the analiticity of $F_{\pi}(s)$ and of
the requirement $min a^F_{\mu} \leq 4.10^{-8}$ makes it possible
to diminish the errors in the values of $F_{\pi}(s_k)$ given in
$^{[1]}$.

In Sec.7 we calculate the values of $F_{\pi}(s_k)$ at 20 additional points
$0.263 GeV^2 \leq s_k \leq 0.463 GeV^2$.

\vspace{5mm}
{\bf 2. Derivation of Minimum $a^F_{\mu}$ Using the Analytical
Properties of $F_{\pi}(s)$ only.}

\vspace{3mm}
Let us map conformally the plane with a cut into interior of the circle $z$
by the formula

%4
\begin{equation}
z = - \frac{\sqrt{s/4m^2_{\pi} - 1} - i}{\sqrt{s/4m^2_{\pi} - 1} + i}
\end{equation}
The value $a^F_{\mu}$ takes the form

%5
\begin{equation}
a^F_{\mu} = \frac{\alpha^2}{12 \pi} ~ \Phi
\end{equation}
where

$$
\Phi = \frac{1}{2 \pi} \int \limits ^{\pi} _{-\pi} f (\theta) \mid F_{\pi}
(e^{i \theta}) \mid^2 ~ d \theta \eqno{(5a)}$$
and

$$
f(\theta) =  K(x)sin^4 \theta/2/cos \theta/2
$$

$$
x = \frac{1 - \sqrt{1 - (m_{\mu}/m_{\pi})^2 cos^2 \theta/2}}{1 + \sqrt{1 -
(m_{\mu}/m_{\pi})^2 cos^2 \theta/2}}
$$
Let us find minimum $\Phi$. Expand the function $F_{\pi}(z)$ in series in
orthogonal polynomials weith the weight function $f(\theta)$:

%6
\begin{equation}
F_{\pi}(z) = \sum _{n=0}^{\infty}~ a_n~P_n(z)
\end{equation}
%7
\begin{equation}
\frac{1}{2 \pi}~ \int \limits_{-\pi} ^{\pi} f(\theta) P^*_n (z) P_m (z)
d\theta = \delta_{nm}, ~~~~~ z = e^{i \theta}
\end{equation}
Substituting expansion (6) into   the formula (5a)
and into the normalization condition   $F_{\pi}(0) = 1$ we
get

%8
\begin{equation}
\Phi = \sum _{n=0} ^{\infty}~ a^2_n
\end{equation}

%eq.9
\begin{equation}
\sum_{n=0} ^{\infty}~ a_n~ P_n(0) = 1
\end{equation}
The coeffiients $a_n$ are real because the function $F_{\pi}(s)$  is real
on the real axis $[-\infty, 4m^2_{\pi}]$.
Minimum $\Phi$ at the additional condition (9) can be found using the
method of Lagrange multipliers.

%10
\begin{equation}
\tilde{\Phi} = \sum_{n=0} ^{\infty}~ a^2_n - \lambda~ \sum_{n=0} ^{\infty}~
a_n~ P_n(0)
\end{equation}
From equation $\frac {\partial \tilde{\Phi}}{\partial a_n} = 0$ we obtain
$a_n = \frac{1}{2} \lambda P_n (0)$ and from eq.(9) it follows\\
$\lambda = \frac{2}{\sum_{n=0} ^{\infty}~ P^2_n(0)}$ and $\Phi_{min} =
 1/\sum_{n=0}^{\infty}~ P^2_n (0)$.

 Let us use the formula from the theory of orthogonal polynomials $^{[3]}$:

 %11
\begin{equation}
A_{km} \equiv \sum_{n=0}^{\infty}~ P^*_n (z_k) P_n(z_m) = \frac{1}{1 -
z_k^* z_m}~~ \frac{1}{D^*(z_k)D(z_m)}
\end{equation}
where
%12
\begin{equation}
D(z) = exp \Biggl \{\frac{1}{4 \pi} \int \limits_{-\infty}^{\infty}~
\frac{1 + ze^{-i \theta}}{1 - ze^{-i \theta}}~ ln f(\theta)d \theta \Biggr \}
\end{equation}
Because of $f(-\theta) = f(\theta)$ we obtain

%13
\begin{equation}
D(z_k) = exp \Biggl \{\frac{1}{2 \pi}~\int \limits_{0}^{\pi}~ \frac{1
 - z^2_k}{1 - 2z_k cos \theta + z^2_k}~ ln f(\theta)~ d\theta \Biggr \}
\end{equation}
and

%14
\begin{equation}
\Phi_{min} = D^2(0) = exp~\Biggl \{\frac{1}{\pi} \int\limits_{0}^{\pi}
ln f (\theta) d \theta \Biggr \}
\end{equation}
After taking the integral (14) we obtain the minimum of $a^F_{\mu}$ following
from analytical properties f.f. $F_{\pi}(s)$ only:

%15
\begin{equation}
\Phi_{min} = 1.13 \cdot 10^{-13},~~~~~~ (a^F_{\mu})_{min} = 1.60 \cdot
10^{-9}
\end{equation}

\vspace{5mm}

{\bf 3. Derivation of Minimum $a^F_{\mu}$ Using the Analytical
Properties}

~~~{\bf of $F_{\pi}(s)$ and the Conditions (3)}

\vspace{3mm}
To derive the minimum of $a^F_{\mu}$ if f.f. $F_{\pi}$  satisfies the
conditions (3) we use the method of Lagrange multipliers:

%16
\begin{equation}
\tilde{\Phi} = \sum_{n=0}^{\infty}~ a^2_n -\sum_{i=0}^{N}~ \lambda_i~
\sum_{n=0} ^{\infty}~ a_n P_n (z_i)
\end{equation}
From condition of minimum $\tilde{\Phi}~  \frac{\partial
\tilde{\Phi}}{\partial a_n} = 0$ we find

%17
\begin{equation}
a_n = \frac{1}{2}~ \sum_{i=1}^{N}~ \lambda_i P_n (z_i)
\end{equation}
The Lagrange multipliers are found from the conditions

%18
\begin{equation}
\frac{1}{2} \sum^{\infty}_{n=0}\lambda_i P_n (z_k) P_n(z_i) = b_i ~~~~~ (i
= 0, 1, ...  N)
\end{equation}
Minimum of $\Phi$ under the conditions (3) is
equal to

%19
\begin{equation}
\Phi_{min} = \frac{1}{2}~ \sum_{i=0}^{N}~ \lambda_i~b_i
\end{equation}
The Lagrange multipliers $\lambda_i$ are found from a system of linear
equations

%20
\begin{equation}
\sum_{k=0}^{N}~ A_{ki}~ \lambda_k = 2b_k,~~~~~ i = 0, 1, ... N
\end{equation}

Substituting $\lambda_k$ from (20) into formula (19) we obtain

%21
\begin{equation}
\Phi_{min} = -\left|
\begin{array}{llll}
0& b_1& ...& b_N\\
b_1& A_{11}&...& A_{1 N}\\
-  & -     & - & - \\
b_N& A_{N 1}& ...& A_{NN}
\end{array}
\right| \cdot \frac{1}{Det (A_{ik})}
\end{equation}
Formula (21) may be simplified by the substitution

%22
\begin{equation}
\tilde{b}_i = b_i~ D(z_i)
\end{equation}

%23
\begin{equation}
\tilde{A}_{ik} = A_{ik} D(z_i)D(z_k)
\end{equation}
$\Phi_{min}$ may be written in new variables in the form

%24
\begin{equation}
\Phi_{min} = \sum_{i, k=0} ^{N}~ C_{ik}~ \tilde{b}_i~ \tilde{b}_k
\end{equation}
where

%25
\begin{equation}
C_{ik} = (-1)^{i+k}~ \tilde{\Delta}_{ik}/Det (\tilde{A}_{ik})
\end{equation}
$\tilde{\Delta}_{ik}$ is the $(i, k)$ minor of the matrix $\tilde{A}_{ik}$.

Det $(\tilde{A}_{ik})$ may be transformed into the form

%26
\begin{equation}
Det(\tilde{A}_{ik}) = (-1)^{i+k}
\frac{(1-z_iz_k)}{(1-z^2_i)(1-z^2_k)}\prod^{N}_{m=0}~^{\prime}~
\frac{z_i-z_m}{1-z_i z_m}~ \prod^{N}_{l=0}~^{\prime}~ \frac{z_k-z_l}{1-
z_kz_l}~ \tilde{\Delta}_{ik}
\end{equation}
Prime above the products implies
that the factors with $m=i$ and $l=k$ should be omitted. As a result we get

%27
\begin{equation}
C_{ik} = \frac{1}{1 - z_i z_k} ~~~ f_i f_k
\end{equation}

%28
\begin{equation}
f_i = (1 - z^2_i)~ \prod^{N}_{m=0}~^{\prime}~ \frac{1 -z_i z_m}{z_i - z_m}
\end{equation}
For $\mid z_i \mid \ll 1$ formula (24) can be transformed into the following
form, which is more convenient  for numerical calculation:

%29
\begin{equation}
\Phi_{min} = \sum^{\infty}_{k=0}~ (\sum^{N}_{i=1}~ f_i \tilde{b}_i z^k_i)^2
\end{equation}

Table I presents the values $Q_k$ in which f.f. $F_{\pi}(Q^2_k)$ is
 measured$^{[1]}$, the values of f.f. at these points, the values of $z_n,
 f_n$ and of min $a^F_{\mu}(n)$  if $F_{\pi}(z_i)$ are equal to the
 central values f.f.  at the points $Q^2_i, i = 0,1,... n$). It is seen
 from Table 1 that even min $a^F_{\mu} (3) > 4.10^{-8}$. If we will use all
 46 points , min $a^h_{\mu}(46)$ will be larger than expected one by
 \underline{a  factor of $10^{100}$.}

\newpage
{\bf 4. Resolution of the Contradiction}

 \vspace{3mm}
 The problem under investigation belongs to the type of the so-called
 "incorrect" problems. The reason is that $\Phi_{min}$ has gigantic
 sensitivity to the measured values of f.f. $F_{\pi}(Q^2_k)$. The
 values $f_i$ have the order of $10^{36}\sim 10^{52}$ but all the sum in
 $\Phi_{min}$ (29) has the order $3 \cdot 10^{-3}$. There is a gigantic
 cancellation in the sum of $\Phi_{min}$.

The contradiction may be resolved if we permit $F_{\pi}(s_k)$ to vary
within the limits of experimental errors. Let us write $\Phi_{min}$ in the
form

%30
\begin{equation}
\Phi_{min} = \sum^{N}_{i, k=0}~ \frac{1}{1 - z_i z_k} x_i x_k
\end{equation}
where $x_i = f_i~ D(z_i) F_{\pi} (z_i), ~~ i = 1,2,...N, ~~~ x_0 =
f_0 D(0)$.

The value $x_0$ is known exactly, the rest of $x_i$ varies so that the
values $F_{\pi}(s_i)$ does not contradict to  experiment.
Let us introduce the functional

%31
\begin{equation}
\tilde{\Phi}_{min} = x^2_0 + 2x_0 \sum^{N-1}_{i=1}~ x_i +
\sum^{N-1}_{i, k=1}~ \frac{1}{1-z_i z_k}~ x_i x_k +
C\sum^{N-1}_{i=1}~ \frac{(x_i-\bar{x}_i)^2}{(\Delta x_i)^2}
\end{equation}
where $\bar{x}_i = f_i D(z_i) \bar{F}_{\pi}(z_i), ~ \bar{F}_{\pi}$
is the value of f.f. $F_{\pi}(z_i)$ in the central point,  $\Delta
F^2_{\pi}(z_i)$ is the experimental error in $F^2_{\pi}(z_i)$.

The constant factor $C$ is chosen in the way that $\chi^2 =\frac{1}{N-1}
\sum^{N-1}_{i=1} \frac{(x_i-\bar{x}_i)^2}{(\Delta x_i)^2}$ is close to 1.

Let us write the condition of minimum $\tilde{\Phi}_{min}, \frac{\partial
\tilde{\Phi}_{min}}{\partial x_i} = 0$ in the form

%32
\begin{equation}
\sum^{N-1}_{k=1}\frac{1}{1 - z_i z_k}~x_k + a_i x_i = -x_0 + a_i \bar{x}_i
\end{equation}
where $a_i = C/(\Delta x_i)^2$. The value $x_0$ is very large and the values
$a_i$ are very small  and for this reason we will solve first the system of
equations

%33
\begin{equation}
\sum^{N-1}_{k=1} \frac{1}{1 - z_i z_k} x^{(0)}_k = -x_0
\end{equation}
The solution of eqs.(33) is

%34
\begin{equation}
x^{(0)}_k = f_k \cdot D (0)
\end{equation}
This solution corresponds to f.f. $F_{\pi}(z)$ minimizing $\Phi$ at one
condition $F_{\pi}(0) = 1$, it corresponds to the f.f. $F(z) =
D(0)/D(z)$. Due to this we make the substitution

%35
\begin{equation}
x_k = x^{(0)}_k + y_k
\end{equation}
For $y_k$ we get the system of linear equations:

%36
\begin{equation}
\sum^{N-1}_{k=1} \frac{1}{1 - z_i z_k}y_k + a_i y_i = a_i(\bar{x}_i -
x^{(0)}_i), ~~~ i = 1,2,...N-1
\end{equation}
The system of equations was solved numerically.
Taking into account that $x_k$ is the solution of
the system of eqs.(32) we obtain

%37
\begin{equation}
min(a^h_{\mu})^{var} = \frac{\alpha^2}{12 \pi^2} \Biggl \{x_0 \Biggl
(\sum^{N-1}_{i=1} x_i + x_0 \Biggr ) + \sum^{N-1}_{i=1} a_i (\bar{x}_i -
x_i) x_i \Biggr \} = 1.02.10^{-8}
\end{equation}

\vspace{5mm}
{\bf 5. Refinements of the values of f.f. obtained in $^{[1]}$}

\vspace{3mm}
Let us fix one of the 45 values of f.f. (e.g., $F_{\pi}(s_k)$) within the
limits of experimental errors. The rest values of f.f. will be varied within
the limits of experimental errors (eqs.(31)--(36)). Let us scan
$F^2_{\pi}(Q_l)$ and then vary $F^2_{\pi}(Q_l) ~~ l \not= k$
until min $a^h_{\mu}$ becomes $4\cdot 10^{-8}$. The constant $C$ changes in
a way that $\chi^2 = 1$.

As a result we obtain the refined values of f.f.
$-(F_{\pi}^2(Q^2)_{ref.}$ (see Table 1).

\vspace{5mm}
{\bf 6. Calculation of the Values $F^2_{\pi}(Q_k)$ at the additional points}

~~~{\bf $Q^2_k = 0.263+0.01 \cdot (k - N),~ N+1, ..., N+20$}

\vspace{3mm}
Let us scan $F^2_{\pi}(Q_k),~ k=N+1,...N+20$ and vary $F^2(Q^2_k),
k=1,...N-1$,
so that $\chi^2 = 1$ and min$a^F_{\mu} \leq
4.10^{-8}$. As a result, we obtain $F^2(Q_k)~~ (k = N+1, ... N+20)$ ~(see
Table 2).

\vspace{5mm}
{\bf 7. Conclusions}

\vspace{3mm}
In conclusion we make some additional comments:

1) There is very large correlation between the values of f.f. at different
$Q_k$. As a rule, if we fix two values $F_{\pi}(Q_k)$ and $F(Q_i)$ within the
limits of experimental errors and will vary the rest of the $F_{\pi}(Q_e)$
values (eqs.(31)-(36)), then the min$a^F_{\mu}$ will be by several orders
of magnitude larger than $4.10^{-8}$.  Thus, the use of the analytical
properties of f.f.  in analysis of the experiments on f.f. measurements
makes it possible to improve significantly the accuracy of the experimental
results.

2) In many theoretical works calculations are made in Euclidean
space. One of the results of this work is that the transition
from Euclidean space to Minkowski space is a nontrivial
problem.

I thank V.L.Morgunov and V.A.Novikov for useful discussions.

%\newpage

%\end{document}
\newpage

\hfill {\bf \Large Table 1}

\bigskip
\begin{center}
\begin{tabular}{|l|l|l|l|l|l|l|}\hline
n & $Q^2$ & $-z_n$ & $(F^2_{\pi}(Q^2))_{Exp}$ & ~~~$f_n$ & min$a^F_{\mu}(n)$
& $(F^2_{\pi}(Q^2))_{ref.}$\\
&&&&&&  \\ \hline
0 & ~0 & ~0 & ~~~1 &  $2.43 \cdot 10^{36}$ & $1.60 \cdot 10^{-9}$ & \\
\hline
1 & 0.015 & 0.044 & 0.944 $\pm$ 0.007 & $-1.25\cdot 10^{45}$ & $2.84\cdot
10^{-9}$ & \\ \hline
2 & 0.017 & 0.0493 & $0.921 \pm 0.006$ & $4.61 \cdot 10^{46}$ & $2.07 \cdot
10^{-6}$ & $0.936 \pm 0.007$ \\ \hline
3 & 0.019 & 0.0545 & $0.933 \pm 0.006$ & $-8.15 \cdot 10^{47}$ & 0.0803 &
$0.932 \pm 0.005$ \\ \hline
4 & 0.021 & 0.0596 & $0.926 \pm 0.006$ & $9.12 \cdot 10^{48}$ & 805.02 & \\
\hline
5 & 0.023 & 0.0646 & $0.914 \pm 0.007$& $-7.22 \cdot 10^{49}$& $3.13 \cdot
10^6$ & \\ \hline
6 & 0.025 & 0.0695 & $0.905 \pm 0.007$ & $4.28 \cdot 10^{50}$ & $5.98 \cdot
10^9$ & \\ \hline
7 & 0.027 & 0.0742 & $0.898 \pm 0.008$ & $-1.96 \cdot 10^{51}$ & $6.32 \cdot
10^{12}$ & \\ \hline
8 & 0.029 & 0.0789 & $0.884 \pm 0.008$ & $7.03 \cdot 10^{51}$ & $4.73 \cdot
10^{15}$ & $0.899 \pm 0.003$ \\ \hline
9 & 0.031 & 0.0835 & $0.884 \pm 0.009$ & $-1.99 \cdot 10^{52}$& $6.74 \cdot
10^{18}$ & $0.885 \pm 0.008$\\ \hline
10 & 0.033 & 0.0881 & $0.890 \pm 0.009$ & $4.44 \cdot 10^{52}$ & $2.66 \cdot
10^{22}$ & $0.884 \pm 0.003$ \\ \hline
11 & 0.035 & 0.0925 & $0.866 \pm 0.01$ & $-7.58 \cdot 10^{52}$& $9.22 \cdot
10^{25}$ & $0.872 \pm 0.004$ \\ \hline
12 & 0.037 & 0.0968 & $0.876 \pm 0.011 $ & $9.43 \cdot 10^{52}$ & $1.62
\cdot 10^{29}$ & $0.869 \pm 0.004$\\ \hline
13 & 0.039 & 0.101 & $0.857 \pm 0.011$ & $-7.26 \cdot 10^{52}$ & $9.26 \cdot
10^{31}$ & $0.861 \pm 0.007$ \\ \hline
14 & 0.042 & 0.107 & $0.849 \pm 0.009$ & $3.80 \cdot 10^{52} $ & $1.45 \cdot
10^{33}$ & $0.853 \pm 0.005$\\ \hline
15 & 0.046 & 0.115 & $0.837 \pm 0.009$ & $-2.71 \cdot 10^{52}$ & $1.56 \cdot
10^{38}$ & $0.841 \pm 0.005$ \\ \hline
16 & 0.050 & 0.123 & $0.83 \pm 0.01$ & $2.96 \cdot 10^{52}$ & $3.22\cdot
10^{41}$ & $0.829 \pm 0.005$ \\ \hline
17 & 0.054 & 0.131 & $0.801 \pm 0.011$ & $-3.65 \cdot 10^{52}$ & $2.88 \cdot
10^{44}$ & $0.818 \pm 0.006$ \\ \hline
18 &0.058 & 0.138 & $0.800 \pm 0.012$ & $4.54 \cdot 10^{52}$& $1.57 \cdot
10^{47}$ & $0.807 \pm 0.005$\\ \hline
19 & 0.062 & 0.145 & $0.809 \pm 0.012$ & $-5.39 \cdot 10^{52}$ & $6.09
\cdot 10^{49}$ & $0.800 \pm 0.003$ \\ \hline
20 & 0.066 & 0.152 & $0.785 \pm 0.014$ & $5.85 \cdot 10^{52}$ & $1.78 \cdot
10^{52}$& $0.785 \pm 0.006$ \\ \hline
21 & 0.070 & 0.159 & $0.785 \pm 0.015$ & $-5.59 \cdot 10^{52}$ & $4.11 \cdot
10^{54}$ & $0.775 \pm0.005$ \\ \hline
22 & 0.074 & 0.165 & $0.777\pm 0.016$ & $4.46 \cdot 10^{52}$ &
$7.79\cdot 10^{56}$ & $0.766 \pm 0.005$  \\ \hline
23 & 0.078 & 0.172 &
$0.769 \pm 0.017$ & $-2.64 \cdot 10^{52}$ & $1.24 \cdot 10^{59}$ & $0.757
\pm 0.005$\\ \hline
24 & 0.083 & 0.179 & $0.757 \pm 0.01$ & $1.13 \cdot
10^{52}$ & $1.63 \cdot 10^{61}$ & $0.746 \pm 0.004$ \\ \hline
25 & 0.089 &
0.188 & $0.715 \pm 0.016$ & $-5.15 \cdot 10^{52}$& $1.73 \cdot 10^{63}$ &
$0.727 \pm0.004$ \\
\hline
26 & 0.095 & 0.197 & $0.724 \pm 0.018$ & $2.97
\cdot 10^{51}$ & $1.61 \cdot 10^{65}$ & $0.715 \pm 0.007$ \\ \hline
27 &
0.101 & 0.205 & $0.680 \pm 0.017$ & $-1.81 \cdot 10^{51}$ & $1.25 \cdot
10^{67}$ & $0.704 \pm 0.008$ \\
\hline
28 & 0.107 & 0.213 & $0.696 \pm 0.019$
& $1.10 \cdot 10^{50}$ & $8.49 \cdot 10^{68}$ & $0.691 \pm 0.009$  \\ \hline
29 & 0.113 & 0.220 & $0.688 \pm 0.020$ & $-6.30 \cdot 10^{50}$ & $5.09 \cdot
10^{70}$ & $0.679 \pm 0.009$ \\ \hline
30 & 0.119 & 0.228 & $ 0.676 \pm
0.021$ & $3.31 \cdot 10^{51}$ & $2.72 \cdot 10^{72}$ & $0.667 \pm 0.009$ \\
\hline
31 & 0.125 & 0.235 & $0.665 \pm 0.023$& $-1.53 \cdot 10^{50}$ & $1.31 \cdot
10^{74}$ & $0.656 \pm 0.009$ \\ \hline
32 & 0.131 & 0.242 & $0.651 \pm 0.024$ & $5.28 \cdot 10^{49}$ & $5.74 \cdot
10^{75}$ & $0.644 \pm 0.010$ \\ \hline
33 & 0.137 & 0.248 & $0.646 \pm 0.027$ & $-1.76 \cdot 10^{49}$ & $2.3 \cdot
10^{77}$ & $0.634 \pm 0.010 $\\ \hline
34 & 0.144 & 0.256 & $0.616 \pm 0.023$ & $3.22 \cdot 10^{48}$ & $8.33 \cdot
10^{78}$ &  $0.621 \pm 0.011$ \\ \hline
35 & 0.153 & 0.265 & $0.654 \pm 0.023$ & $-3.97 \cdot 10^{47}$ & $2.69 \cdot
10^{80}$ & $0.605 \pm 0.013$ \\
&&&&&& \\ \hline
\end{tabular}
\end{center}

\newpage
\begin{center}
\begin{tabular}{|l|l|l|l|l|l|l|}\hline
&&&&&&\\
36 & 0.163 & 0.275 & $0.563 \pm 0.024 $ & $5.42 \cdot 10^{46}$ & $7.73 \cdot
10^{81}$  & $0.580 \pm 0.007$ \\ \hline
37 & 0.173 & 0.284 & $0.534 \pm 0.038$ & $-9.38 \cdot 10^{45}$ & $2 \cdot
10^{83}$ & $ 0.565 \pm 0.009$ \\ \hline
38 & 0.183 & 0.293 & $0.586 \pm 0.034$ & $1.69 \cdot 10^{45}$ & $4.66 \cdot
10^{84}$  & $0.564 \pm 0.012$ \\ \hline
39 & 0.193 & 0.302 & $0.544 \pm 0.036$ & $-2.95 \cdot 10^{44}$ & $9.91 \cdot
10^{85}$ & $0.544 \pm 0.020$ \\ \hline
40 &  0.203 & 0.310 & $0.529 \pm 0.040$ & $4.74 \cdot 10^{43}$ & $1.93 \cdot
10^{87}$  & $0.528 \pm 0.024$ \\ \hline
41 & 0.213 & 0.318 & $0.616 \pm 0.048$ & $-6.65 \cdot 10^{42}$ & $3.45 \cdot
10^{88}$ & $0.512 \pm 0.030$ \\ \hline
42 & 0.223 & 0.326 & $0.487 \pm 0.049$ & $7.70 \cdot 10^{41}$ & $5.69 \cdot
10^{89}$ & $0.500\pm 0.032$\\ \hline
43 & 0.233 & 0.333 & $0.417 \pm 0.058$ & $-6.82 \cdot 10^{40}$ & $8.69 \cdot
10^{91}$ & $0.465 \pm 0.030$ \\ \hline
44 & 0.243 & 0.340 & $0.593 \pm 0.074$ & $4.07 \cdot 10^{39}$ & $1.23 \cdot
10^{92}$ & $0.470 \pm 0.035$ \\ \hline
45 & 0.253 & 0.347 & $0.336 \pm 0.074$ & $-1.22 \cdot 10^{38}$ & $1.65 \cdot
10^{93}$  & $0.458 \pm 0.040$\\
&&&&&& \\ \hline
\end{tabular}
\end{center}

\vspace{1.5cm}
\hspace{10cm} {\bf \large Table 2}

\vspace{5mm}
\begin{center}
\begin{tabular}{|l|l|l|l|l|l|}\hline
$n$ & ~~$Q^2$ & ~~$F^2_{\pi}(Q^2)$ & $n$ & ~~$Q^2$ & ~~$F^2_{\pi}(Q^2)$\\
&&&&& \\ \hline
1 & 0.263 & $0.448 \pm 0.043 $ & 11 & 0.363 & $0.357 \pm 0.077$\\ \hline
2 & 0.273 & $0.439 \pm 0.046$ & 12 & 0.373 & $0.349 \pm 0.082$\\ \hline
3 & 0.283 & $0.428 \pm 0.049$ & 13 & 0.383 & $0.342 \pm 0.085$ \\ \hline
4 & 0.293 & $0.417 \pm 0.052$ & 14 & 0.393 & $0.336 \pm 0.086$ \\ \hline
5 & 0.303 & $0.407 \pm 0.056$ & 15 & 0.403 & $0.330 \pm 0.089$ \\ \hline
6 & 0.313 & $0.398\pm 0.060 $ & 16 & 0.413 & $0.324 \pm 0.093$ \\ \hline
7 & 0.323 & $0.389 \pm 0.064$ & 17 & 0.423 & $0.318 \pm 0.096$ \\ \hline
8 & 0.333 & $0.381 \pm 0.067$ & 18 & 0.433 & $0.312 \pm 0.099$ \\ \hline
9 & 0.343 & $0.373 \pm 0.070$ & 19 & 0.443 & $0.306 \pm 0.102$ \\ \hline
10 & 0.353 & $0.365 \pm 0.073$ & 20 & 0.453 & $0.301 \pm 0.105$ \\
&&&&&\\ \hline
\end{tabular}
\end{center}

\begin{thebibliography}{99}
\bibitem{1} B.R.Amendolia et al., Nucl.Phys. B277(1986) 168.
\bibitem{2} M.Gourdin and E.de Rafael, Nucl.Phys. B10 (1969)
667;\\
L.Durand, Phys.Rev. 128(1962) 441.
\bibitem{3} G.Szego, Orthogonal polynamial 1959.
\end{thebibliography}
\end{document}